\documentclass[runningheads]{llncs}
\usepackage[T1]{fontenc}
\usepackage{graphicx}
\graphicspath{{images/}}
\usepackage{amssymb}
\usepackage{physics}
\usepackage{algorithm}
\usepackage{algpseudocode}

\newcommand{\Beginproof}{{\em Proof.}  }
\newcommand{\Endproof}{\hfill$\Box$\\}

\begin{document}
\title{Quantum Algorithm for Searching for the Longest Segment\\ and the Largest Empty Rectangle}
%
%
\author{Kamil Khadiev\inst{1,4}\orcidID{0000-0002-5151-9908}\and Vladislav Remidovskii\inst{2} \and Timur Bikmullin\inst{1} \and Aliya Khadieva\inst{1,3,4}}
\authorrunning{K. Khadiev et al.}
%
\institute{Institute of Computational Mathematics and Information Technologies, Kazan Federal University, Kazan, Tatarstan, Russia \and
Yandex LTD., Moscow, Russia
\and
University of Latvia, Riga, Latvia
\and Zavoisky Physical-Technical Institute, FRC Kazan Scientific Center of RAS, Kazan, Russia\\
\email{kamilhadi@gmail.com}}
\maketitle              
\begin{abstract}
In the paper, we consider the problem of searching for the Largest empty rectangle in a 2D map, and the one-dimensional version of the problem is the problem of searching for the largest empty segment. We present a quantum algorithm for the Largest Empty Square problem and the Largest Empty Rectangle of a fixed width $d$ for $n\times n$-rectangular map. Query complexity of the algorithm is $\tilde{O}(n^{1.5})$ for the square case, and $\tilde{O}(n\sqrt{d})$ for the rectangle with a fixed width $d$ case, respectively. At the same time, the lower bounds for the classical case are $\Omega(n^2)$, and $\Omega(nd)$, respectively. The Quantum algorithm for the one-dimensional version of the problem has $O(\sqrt{n}\log n\log\log n)$ query complexity. The quantum lower bound for the problem is $\Omega(\sqrt{n})$ which is almost equal to the upper bound up to a log factor. The classical lower bound is $\Omega(n)$. So, we obtain the quadratic speed-up for the problem. 

\end{abstract}
%
%
%
\section{Introduction}
\label{sec:intro}
Quantum computing \cite{nc2010} is one of the hot topics in computer science of the last decades.
There are many problems where quantum algorithms outperform the best known classical ones \cite{dw2001,quantumzoo},
and one of the most important performance metrics in this regard is query complexity.
We refer to \cite{a2017,aazksw2019part1,k2022lecturenotes} for a nice survey on the quantum query complexity,
and to \cite{aaksv2022,asa2024,aj2021,kkmsy2022,ki2019,kbcw2024,ke2022,kiv2022,kk2021,kb2022,kr2021b,kr2021a,kszm2022,ks2025,l2020,l2020conf,m2017,ks2024,asa2025,ks2019,kks2019,kksk2020,gnbk2021,ks2023} for the more recent progress on string processing and graph problems.

In the paper, we consider The Largest Empty Rectangle problem and its one-dimensional version The Longest Empty Segment Problem. In the one-dimensional case, we want to search for the longest segment of elements that do not satisfy some predicate among a sequence of $n$ elements. Informally, we call such a segment ``empty''. In the two-dimensional case, it is a largest ``empty'' rectangle in $n\times n$-rectangular map.  These problems have many applications including ``maximal white rectangles'' in the image segmentation R and D of image processing and pattern recognition \cite{baird1990image};   electronic design automation, design and verification of physical layout of integrated circuits \cite{ullma1984computational}.
Typically, researchers consider restricted versions of the problem like points as obstacles \cite{naamad1984maximum,chan2023faster,bae2023empty} or line segment obstacles  \cite{nandy1990efficient,nandy1994location}. Here, we consider a more general case.

In the one-dimensional case, we present a quantum algorithm with $O(\sqrt{n}\log n\log\log n)$ query complexity. We show that it is equal to the lower bound up to the log factor. We prove the quantum lower bound $\Omega(\sqrt{n})$. At the same time, we reach almost quantum quadratic speed-up. The lower bound in the classical (deterministic or randomized) case is $\Omega(n)$. The one-dimensional case has many applications also including searching for data with a specific property in a data sequence and others. 

In the two-dimensional case, we reach not so significant, but still important speed-up. We present quantum solutions for restricted versions of the problem that are The Largest Empty Rectangle Problem with a Fixed Width $d$ ($LREC2$), The Largest Empty Square Problem($LSQR$), and The Largest Empty Rectangle Problem for Rectangle Empty Areas ($LRECW$) problems. The query complexity is $O(n^{1.5})$ for $LREC2$, is $O(n^{1.5}\log n\log \log n)$ for $LSQR$, and is $O(n\sqrt{d}\log n\log \log n)$ for $LRECW$.
At the same time, the lower bounds for classical (deterministic or randomized) solutions are $\Omega(n^2)$, $\Omega(n^2)$, and $\Omega(nd)$, respectively. So, we obtain quantum polynomial speed-up.
At the same time, the presented quantum lower bounds are $\Omega(n)$, $\Omega(n)$, and $\Omega(\sqrt{nd})$, respectively. That is far from the upper bounds. Additionally, we presented the lower bound of The Largest Empty Rectangle problem in the general case, that is $\Omega(n)$ in the quantum case, and $\Omega(n^2)$ in the classical case.

\textbf{Comparing with existing results.}
Note that the one-dimensional problem is close to the pattern matching problem (searching for a substring in a text) which quantum solution was presented in \cite{rv2003}. At the same time, in our problem we are not supposed to search for the exact pattern with a fixed length, but the segment with the largest length with some property. So, we cannot apply the algorithm from \cite{rv2003} directly. A similar situation is in the two-dimensional case and the quantum algorithm from \cite{m2017}.

There is an interesting quantum algorithm for the longest palindrome substring (LPS) problem \cite{l2020}, and a modification of this algorithm for the longest square substring (LSS) problem \cite{aj2021}. These algorithms solve similar problems and use similar techniques like Grover's search \cite{g96,bbht98} and amplitude amplification \cite{bhmt2002}. At the same time, the The Longest Empty Segment Problem cannot be presented as a partial case of the LPS problem as it was done with LSS problem; and the algorithm for LPS cannot be directly converted for The Longest Empty Segment Problem.

Researchers \cite{ags2019,ckksw2022} are interested in quantum query complexity of recognition of star-free languages. One of the questions is whether a string $s\in\{0,1,2\}^*$ of a length $n$ contains a substring of the form $20^*2$, i.e., two $2$s on both sides of an arbitrarily long string of $0$s. The problem can be easily solved using the modification of our algorithm for The Longest Empty Segment Problem. We present an algorithm with $O(\sqrt{n\log n})$ query complexity. The complexity is the same as in papers  \cite{ags2019,ckksw2022}. At the same time, our technique is much more simple comparing with the quantum query complexity trichotomy \cite{ags2019} and quantum  divide and conquer \cite{ckksw2022} techniques.

Note that the the Longest Empty Segment Problem (one-dimensional version of the problem) was independently discovered in \cite{abbls2025}. There researchers solved the problem using quantum version of the divide and conquer. In the current paper we use much more simple technique based on Grover's search \cite{g96,bbht98} and amplitude amplification \cite{bhmt2002}.

\textbf{The structure of the paper.}
 Preliminaries are situated in Section \ref{sec:prelims}. We discuss the one-dimensional case in Section \ref{sec:lseg}, and the two-dimensional case in Section \ref{sec:lrec}. Section \ref{sec:conclusion} contains the conclusions.

\section{Preliminaries}
\label{sec:prelims}
Firstly, let us define the one-dimensional problem.

{\bf The Largest Empty Segment Problem.}
Let us consider a function $f:\{0,\dots,n-1\}\to \{0,1\}$ for some integer $n$. For $0\leq l\leq r\leq n-1$, let $(l,r)$ be {\em an empty segment} if $f(i)=0$ for any $l\leq i\leq r$. Let a length of an empty segment be $len(l,r)=r-l+1$. Our goal is to search for an empty segment with the largest length. Formally, our goal is to find an empty segment $(l,r)$ such that there is no other empty segment $(l',r')$ such that $r'-l'+1>r-l+1$.
Let us call the problem $LSEG(f,n)$.



There are several ways to generalize the problem for the two-dimensional case. Let us present the first one which is the square case.

{\bf The Largest Empty Square Problem.} 
Let us consider a function $f:\{0,\dots,n-1\}\times\{0,\dots,n-1\}\to \{0,1\}$ for some integer $n$. For $0\leq x\leq n-1, 0\leq y\leq n-1$ and $d:d\leq n$, let $(x,y,d)$ be {\em an empty square} if $f(i,j)=0$ for any $x\leq i\leq x+d-1$, $y\leq j\leq y+d-1$ and $d\leq n-x, d\leq n-y$. Let $d$ be the size of the square. Our goal is to search for an empty square with the largest size. Formally, our goal is to find an empty square $(x,y,d)$ such that there is no other empty square $(x',y',d')$ such that $d'>d$. Let us call the problem $LSQR(f,n)$.

The rectangle case is as follows.

{\bf The Largest Empty Rectangle Problem.} 
Let us consider a function $f:\{0,\dots,n-1\}\times\{0,\dots,n-1\}\to \{0,1\}$ for some integer $n$. Let us have two integer parameters $h$ and $w$. For $0\leq x_1\leq x_2\leq n-1, 0\leq y_1\leq y_2\leq n-1$, let $(x_1,y_1,x_2,y_2)$ be {\em an empty rectangle} if $f(i,j)=0$ for any $x_1\leq i\leq x_2$, $y_1\leq j\leq y_2$ and $y_2-y_1+1\geq h, x_2-x_1+1\geq w$. Let $size(x_1,y_1,x_2,y_2)=(x_2-x_1)\cdot(y_2-y_1)$ be a size (area) of the rectangle. Our goal is to search for an empty rectangle with the largest size. Formally, our goal is to find an empty rectangle $(x_1,y_1,x_2,y_2)$ such that there is no other empty rectangle $(x_1',y_1',x_1',y_2')$ such that $size(x_1',y_1',x_1',y_2')>size(x_1,y_1,x_1,y_2)$. Let us call the problem $LREC(f,n,h,w)$.

Let us present two restricted versions of the problem. The first one is for a fixed width.

{\bf The Largest Empty Rectangle Problem with a Fixed Width.} 
Let us consider a function $f:\{0,\dots,n-1\}\times\{0,\dots,n-1\}\to \{0,1\}$ for some integer $n$. Let us have an integer parameter $d$. Our goal is to search for an empty rectangle of width $d$ with the largest size (area). Formally, our goal is to find an empty rectangle $(x_1,y_1,x_1+d-1,y_2)$ such that there is no other empty rectangle $(x_1',y_1',x_2'+d-1,y_2')$ such that $size(x_1',y_1',x_1'+d-1,y_2')>size(x_1,y_1,x_1+d-1,y_2)$. Let us call the problem $LRECW(f,n,d)$.

The second one is a restriction for empty areas of $f$.

{\bf The Largest Empty Rectangle Problem for Rectangle Empty Areas.} 
Let us consider a function $f:\{0,\dots,n-1\}\times\{0,\dots,n-1\}\to \{0,1\}$ for some integer $n$. Let us have two integer parameters $h,w$. 
Assume that any empty area is a rectangle. Formally, $f(i,j)=0$ iff there is an empty rectangle $(x_1,y_1,x_2,y_2)$ such that $x_1\leq i\leq x_2$ and $y_1\leq j \leq y_2$ and
\begin{itemize}
\item $x_1=0$ or $f(x_1-1,j')=1$ for $y_1\leq j' \leq y_2$;
\item $x_2=n-1$ or $f(x_2+1,j')=1$ for $y_1\leq j' \leq y_2$;
\item $y_1=0$ or $f(i',y_1-1,)=1$ for $x_1\leq i' \leq x_2$;
\item $y_2=n-1$ or $f(i',y_2+1)=1$ for $x_1\leq i' \leq x_2$.
\end{itemize} 
Our goal is to search for an empty rectangle with the largest size such that its height is at least $h$ and its width is at least $w$. Let us call the problem $LREC2(f,n,h,w)$.

\textbf{Quantum query model.}
One of the most popular computation models for quantum algorithms is the query model.
We use the standard form of the quantum query model. 
Let $f:D\rightarrow \{0,1\},D\subseteq \{0,1\}^M$ be an $M$ variable function. Our goal is to compute on an input $x\in D$. We are given an oracle access to the input $x$, i.e. it is implemented by a specific unitary transformation usually defined as $\ket{i}\ket{z}\ket{w}\rightarrow \ket{i}\ket{z+x_i\pmod{2}}\ket{w}$, where the $\ket{i}$ register indicates the index of the variable we are querying, $\ket{z}$ is the output register, and $\ket{w}$ is some auxiliary work-space. An algorithm in the query model consists of alternating applications of arbitrary unitaries which are independent on the input and the query unitary, and a measurement at the end. The smallest number of queries for an algorithm that outputs $f(x)$ with probability $\geq \frac{2}{3}$ on all $x$ is called the quantum query complexity of the function $f$ and is denoted by $Q(f)$.
We refer the readers to \cite{nc2010,a2017,aazksw2019part1,k2022lecturenotes} for more details on quantum computing. 
In this paper, we are interested in the query complexity of the quantum algorithms. We use modifications of Grover's search algorithm \cite{g96,bbht98} as quantum subroutines. For these subroutines, time complexity can be obtained from query complexity by multiplication to a log factor \cite{ad2017,g2002}.   

\section{1D: the Largest Empty Segment Problem and Searching for a Segment of 0s bounded by 2s}\label{sec:lseg}
In this section we consider two kind of  one-dimensional problems that are the Largest Empty Segment Problem in Section \ref{sec:lsegm}, and Searching for a Segment of 0s bounded by 2s in Section \ref{sec:twos}.
\subsection{The Largest Empty Segment Problem}\label{sec:lsegm}
Let us start from the Largest Empty Segment Problem $LSEG(f,n)$.
Firstly, we solve a simple version of the problem. Let us fix a point $i$ such that $0\leq i\leq n-1$; and a size $d$ such that $1\leq d\leq n$. We want to implement a subroutine $\textsc{FixedLenFixedPoint}(f,i,d)$ such that if $i$ belongs to some empty segment of the size $d$, then the subroutine returns the segment; and the subroutine returns $(NULL,NULL)$ otherwise. Formally, the subroutine returns some $(l,r)$ such that $0\leq l\leq i\leq r\leq n$, $(l,r)$ is an empty segment and $len(l,r)=d$.
For the subroutine implementation, we use a quantum algorithm that is a modification of the Grover search algorithm \cite{g96,bbht98}. It finds the minimal index $i$ from the segment $(L,R)$ such that $f(i)=1$. Different implementations of the algorithm are discussed in \cite{dhhm2006,k2014,ll2015,ll2016,kkmsy2022}.
\begin{lemma}\label{lm:firstone}
For a function $f$ and two indices $0\leq L\leq R\leq n-1$, the algorithm searches for the minimal index $i$ such that $f(i)=1$ and $L\leq i\leq R$. If there is no such index, then it returns $NULL$. The query complexity of the algorithm is $O(\sqrt{i-L}\cdot T_f)$ if the one exists; and $O(\sqrt{R-L}\cdot T_f)$ if there is no one in the segment. The error probability is at most $0.1$. Here, $T_f$ is the query complexity of computing $f$.
\end{lemma}
Assume that we have $\textsc{FirstOne}(f,L,R)$ subroutine that implements the algorithm from the previous lemma.
Assume that we have a similar function $\textsc{LastOne}(f,L,R)$ subroutine that searches for the maximal index $i$ such that $f(i)=1$ and $L\leq i\leq R$ or returns $NULL$ if there is no such index. We can implement it using the same algorithm.
The idea of an implementation of the subroutine $\textsc{FixedLenFixedPoint}(f,i,d)$ is the following. 
    
    {\bf Step 1} We check that $f(i)=0$. If $f(i)=1$, then we return $(NULL,NULL)$ and stop the algorithm.

{\bf Step 2} We search for the closest index of $1$ to the right of $i$ such that the distance between $1$ and $i$-th element is at most $d$. The index can be computed as $r\gets\textsc{FirstOne}(f,i,min(i+d-1,n-1))$. If there is no $1$ in the searching segment (result is $NULL$), then we assign $r\gets min(i+d-1,n-1)$. After that, we go to Step 3.

{\bf Step 3} We do the same, but to the left of $r$. We search index of $1$ and it is $\ell\gets\textsc{LastOne}(f,max(0,r-d+1),r)$. If there is no $1$ in the searching segment (result is $NULL$), then we assign $\ell\gets max(0,r-d+1)$. 
     If $r-\ell+1<d$, then $i$ does not belong to the empty segment of the size $d$, and we return $(NULL,NULL)$. Otherwise, we return $(\ell, r)$.
The implementation is presented in Algorithm \ref{alg:fixlen} (Appendix \ref{apx:alg-fixlen}), and Lemma \ref{lm:cmpl-fixlen} describes the complexity.
   
\begin{lemma}\label{lm:cmpl-fixlen}
The query complexity of Algorithm \ref{alg:fixlen} for $\textsc{FixedLenFixedPoint}(f,i,d)$ subroutine is $O(\sqrt{d}\cdot T_f)$ and the error probability is at most $0.2$, where $T_f$ is the query complexity of computing $f$. 
\end{lemma}
\Beginproof
The query complexity of $\textsc{LastOne}$ and $\textsc{FirstOne}$ is $O(\sqrt{d}\cdot T_f)$. We invoke them sequentially. Therefore, the total query complexity is $O(\sqrt{d}\cdot T_f)$. We have two independent error events. So, the total error probability is at most $0.1\cdot 2=0.2$.
\Endproof
Let us generalize the problem a little. We fix the size $d$ such that $1\leq d\leq n$. We want to implement the subroutine $\textsc{FixedLen}(f,d)$ that returns any empty segment $(l,r)$ of the length $d$ and $(NULL,NULL)$ if there is no such a segment.
We suggest a randomized algorithm that implements the subroutine. Let us uniformly randomly choose $i\in_R\{0,\dots,n-1\}$; invoke $\textsc{FixedLenFixedPoint}(f,i,d)$; and return its result.

Let us discuss the success probability of the implementation. If there is no empty segment of the length $d$, then the randomized implementation always returns $(NULL,NULL)$. If there is at least one such segment, then the probability of the correct answer is the probability of picking an index $i$ inside an empty segment of the length $d$. At the same time, we have the probability of an error for $\textsc{FixedLenFixedPoint}(f,i,d)$. So, the total success probability is $p_{success}\geq\frac{0.8\cdot d'}{n}$ if the maximal length of an empty segment is $d'$ and $d\leq d'$.

We apply the Amplitude Amplification algorithm \cite{bhmt2002} for the randomized implementation and obtain error probability $0.1$. If $d=O(d')$, then the total query complexity of the algorithm is $O(\sqrt{\frac{1}{p_{success}}}\cdot\sqrt{d}\cdot T_f)=O(\sqrt{\frac{n}{d'}}\cdot\sqrt{d}\cdot T_f)=O(\sqrt{n}\cdot T_f)$.
The error probability for the amplitude amplification implementation of  $\textsc{FixedLen}(f,d)$ is $0.1$.

Finally, let us discuss the algorithm for $LSEG(f,n)$ problem.
Let $d'$ be the maximal length of an empty segment. We can say that for any $d\leq d'$, the subroutine $\textsc{FixedLen}(f,d)$ returns a non-null result. At the same time, for any $d>d'$ it returns $(NULL,NULL)$. So, we can use Binary search and Binary lifting-like ideas for searching for an answer. For this propose, we do the following two steps:

    
    {\bf Step 1.} We consider powers of $2$ as possible values of $d$. In other words, we invoke $\textsc{FixedLen}(f,d)$ for $d = 2,4,\dots, 2^j$. We search for the first $j$ such that $\textsc{FixedLen}(f,d)=(NULL,NULL)$. We can be sure that $2^{j-1}\leq d'<2^j$. 

{\bf Step 2.} We invoke the Binary search for $d$ for $[2^{j-1},2^j]$ left and right bounds and find the target $d'$ with the target segment.
Due to $2^{j-1}\leq d'<2^j$, we can claim that $d=O(d')$ for any invocation of $\textsc{FixedLen}$ on both steps.
For obtaining a small probability of error, we repeat each invocation of $\textsc{FixedLen}(f,d)$ subroutine $\log\log n$ times, and check whether one of the invocations returns a non-null result. We can perform it because we have a one-side error for the subroutine.


The implementation is presented in Algorithm \ref{alg:lseg} (Appendix \ref{apx:alg-lseg}), and Theorem \ref{th:compl-lseg} describes the complexity.
\begin{theorem}\label{th:compl-lseg}
Algorithm \ref{alg:lseg} solves $LSEG(f,n)$ and has $O(\sqrt{n}\log n\cdot \log \log n\cdot T_f)$ query complexity and $O\left(\frac{1}{\log n}\right)$ error probability. (See Appendix \ref{apx:compl-lseg})
\end{theorem}
Let us discuss the classical (deterministic or randomized) and quantum lower bounds for the problem's complexity.
\begin{theorem}\label{th:compl-lseg2}
Assume that $T_f=O(1)$, then the lower bound for the query complexity for $LSEG(f,n)$ is $\Omega(n)$ in the classical case, and $\Omega(\sqrt{n})$ in the quantum case. (See Appendix \ref{apx:compl-lseg2})
\end{theorem}
Note that the upper bound and the lower bound are the same up to the log factor.

\subsection{Quantum Algorithm for Searching for a Segment of 0s bounded by 2s}\label{sec:twos}
Let us present the problem formally. Assume that we have a function $f:\{0,\dots,n-1\}\to \{0,1,2\}$ for some integer $n$. For $0\leq l\leq r\leq n-1$, let $(l,r)$ be {\em a segment of $0$s bounded by $2$s} if $f(i)=0$ for any $l< i< r$ and $f(l)=f(r)=2$. Our goal is to search for any segment of $0$s bounded by $2$s. Let us call the problem $SZBT(f,n)$.
%
The general idea of the algorithm is the same as for the previous problem. At the same time, we are not supposed to search for the maximal segment, but we can find any of them. 
Firstly, we solve the simple problem. Let us fix a point $i$ such that $0\leq i\leq n-1$; and a segment's size $d$ is such that $1\leq d\leq n$. We want to implement a subroutine $\textsc{FixedLenFixedPointSZBT}(f,i,d)$ such that if $i$ belongs to a segment of $0$s bounded by $2$s of a size at most $d$, then the subroutine returns it; and the subroutine returns $(NULL,NULL)$ otherwise. 
Let us consider a function  $f':\{0,\dots,n-1\}\to \{0,1\}$ such that $f'(i)=1$ iff $f(i)\neq 0$. This function indicates non-zero elements of $f$ and is used as a search function for $\textsc{FirstOne}$ and $\textsc{LastOne}$ functions. For simplicity of explanation, we assume that $f$ and $f'$ are extended by one element to the left and right, and $f(-1)=f(n)=2$; $f'(-1)=f'(n)=1$ 
The idea of an implementation of the subroutine $\textsc{FixedLenFixedPointSZBT}(f,i,d)$ is the following. 
   
    {\bf Step 1} We check that $f(i)=0$. If $f(i)=1$ or $f(i)=2$, then we return $(NULL,NULL)$ and stop the algorithm.

{\bf Step 2} We search for the closest index of $2$ or $1$ to the right of $i$ such that the distance between $i$ and the target element is at most $d$. The index is $r\gets\textsc{FirstOne}(f',i,min(i+d-1,n))$. If there is no target element (the result is $NULL$), then  we assign $r\gets min(i+d-1,n)$. If $f(r)=1$, then the algorithm is failed because the segment of $0$s is not bounded by $2$s; we return $(NULL,NULL)$ and stop the algorithm. Otherwise, we go to Step 3.  

{\bf Step 3} We search for the closest index of $2$ or $1$ to the left of $r$ such that the distance is at most $d$. The index is  $\ell\gets\textsc{LastOne}(f',r-d+1,r-1)$. If there is no target element (the result is $NULL$) or $f(\ell)=1$, then the algorithm is failed because the segment of $0$s is not bounded by $2$s. In that case, we return $(NULL,NULL)$. Otherwise, we return the $(\ell,r)$ segment.

The implementation is presented in Algorithm \ref{alg:fixlenszbt} (Appendix \ref{apx:alg-fixlenszbt}), and Lemma \ref{lm:cmpl-fixlenszbt} describes the complexity.
\begin{lemma}\label{lm:cmpl-fixlenszbt}
The query complexity of Algorithm \ref{alg:fixlenszbt} for $\textsc{FixedLenFixedPointSZBT}(f,i,d)$ subroutine is $O(\sqrt{d}\cdot T_f)$ and the error probability is at most $0.2$, where $T_f$ is the query complexity of computing $f$.
\end{lemma}
The proof is the same as for Lemma \ref{lm:cmpl-fixlen}.
%
Let us generalize the problem a little. Let us fix the size $d$ such that $1\leq d\leq n$. We want to implement a subroutine $\textsc{FixedLenSZBT}(f,d)$ that returns any segment $(l,r)$ of $0$s bounded by $2$s of a length at most $d$ and $(NULL,NULL)$ if there is no such a segment.
We suggest a randomized algorithm that implements the subroutine. Let us uniformly randomly choose $i\in_R\{0,\dots,n-1\}$; invoke $\textsc{FixedLenFixedPointSZBT}(f,i,d)$; and return its result. 
Similarly, the success probability is $p_{success}\geq\frac{0.8\cdot d''}{n}$ if the minimal length of an empty segment is $d''$ and $d\geq d''$. 
We apply the Amplitude Amplification algorithm \cite{bhmt2002} for the randomized implementation and obtain error probability $0.1$. If $d=O(d')$, then the total query complexity of the algorithm is $O(\sqrt{\frac{1}{p_{success}}}\cdot\sqrt{d}\cdot T_f)=O(\sqrt{\frac{n}{d'}}\cdot\sqrt{d}\cdot T_f)=O(\sqrt{n}\cdot T_f)$.
The error probability for the amplitude amplification implementation of  $\textsc{FixedLenSZBT}(f,d)$ is $0.1$.

Finally, let us discuss the algorithm for the main problem.
Let $d''$ be the minimal length of an empty segment. We can say that for any $d\geq d'$, the subroutine $\textsc{FixedLen}(f,d)$ returns a non-null result. At the same time, for any $d<d'$ it returns $(NULL,NULL)$. 
Let us consider the sequence of all powers of $2$ from $1$ up to $n$ as possible values for $d$ that is ${\cal D}=\{2,4,8,\dots 2^{\lceil\log_2 n\rceil},n\}$. So, we want to find the minimal $j$ such that for $d=2^j$ we find the segment or say that there is no such a segment if we no not find the required element $j$. In that case, $2^{j-1} \leq d'\leq 2^{j}$ and $d=O(d')$ for all invocations. We can use First One Search idea for this search problem also and find the required element in $O(\sqrt{d'})=O(\sqrt{|{\cal D}|})=O(\sqrt{\log n})$ steps. Note that here, we have nested Grover algorithms invocations and we should use the bounded-error input version of the algorithm \cite{hmw2003,abikkpssv2020,abikkpssv2023}. Finally, we have the following complexity of the problem.
\begin{theorem}\label{th:compl-lsegbtz}
The presented algorithm solves $STBZ(f,n)$ and has $O(\sqrt{n\log n}\cdot T_f)$ query complexity and at most $0.1$ error probability. If $T_f=O(1)$, then the query complexity is $O(\sqrt{n\log n})$.
\end{theorem}
\Beginproof The complexity of $\textsc{FixedLenSZBT}$ is $O(\sqrt{d'}\cdot T_f)$. The complexity of outer First One Search algorithm is $O(\sqrt{\log d'}\sqrt{d'}\cdot T_f)=O(\sqrt{n\log n}\cdot T_f)$. The error probability is correct because of properties of First One Search algorithm (Lemma \ref{lm:firstone}) and the properties of the bounded-error input version of the Grover's search algorithm \cite{hmw2003,abikkpssv2020,abikkpssv2023}.
\Endproof
\section{2D: The Largest Empty Square Problem and the Largest Empty Rectangle Problems}\label{sec:lrec}
Let us consider the two-dimensional case.

\textbf{The Largest Empty Square Problem. }
Let us start from the $LSQR(f,n)$ problem.
We use a solution similar to the one-dimensional case.
Firstly, we solve a simple problem. Let us fix a point $(i,j)$ such that $0\leq i,j\leq n-1$; and a size $d$ such that $1\leq d\leq n$. We want to implement a subroutine $\textsc{FixedSizeFixedPointSquare}(f,i,j,d)$ such that if $(i,j)$ belongs to an empty square of the size $d$, then the subroutine returns it; and the subroutine returns $(NULL,NULL,NULL)$ otherwise. Formally, the subroutine returns $(x,y,d)$ such that $0\leq x\leq i\leq x+d-1$,  $0\leq y\leq j\leq y+d-1$, and $(x,y,d)$ is an empty square.
In the two dimensional case, we can use $f(i)$ as a $f(i):\{0,\dots,n-1\}\to\{0,1\}$ fucntion where $0\leq i\leq n-1$ and $f(i)(j)=f(i,j)$.
Here, we use a modification of $\textsc{FirstOne}(f(i),L,R)$ subroutine, that searches for the {\bf maximal} index $j\in\{L,\dots,R\}$ such that $f(i,j)=1$. Let us call it $\textsc{LastOne}(f(i),L,R)$. It has the same properties as the original subroutine \cite{dhhm2006,k2014,ll2015,ll2016,kkmsy2022}. 
The idea of an implementation of the subroutine $\textsc{FixedSizeFixedPointSquare}(f,i,j,d)$ is the following. 
   
   {\bf Step 1} We check that $f(i,j)=0$. If $f(i,j)=1$, then we return $(NULL,NULL,NULL)$ and stop the algorithm.

{\bf Step 2} For each $k$ such that $i-d+1\leq k \leq i+d-1$ we compute two numbers $l_k$ and $r_k$. The value $l_k$ is the closest $1$ in the $k$-row to the index $j$ in the left direction. The value $r_k$ is similar but in the right direction. We can compute them as $l_k\gets\textsc{LastOne}(f(k),j,j-d+1)$. If $\textsc{LastOne}(f(k),j,j-d+1)=NULL$, then we set $l_k\gets j-d$. We compute $r_k\gets\textsc{FirstOne}(f(k),j,j+d-1)$.
     If $\textsc{FirstOne}(f(k),j,j+d-1)=NULL$, then we set $r_k\gets j+d$.   Note that the step is a sequence of invocations of the First One Search algorithm. Each invocation can have an error and the total error accumulates very fast. According to \cite{k2014}, it can be converted to the algorithm that does the same action, has the same complexity but has a total error $0.1$.

{\bf Step 3} For each $y\in \{j-d+1,\dots,j\}$ we check $l'(y)=\max\limits_{y\leq k\leq y+d-1}l_k$ and $r'(y)=\min\limits_{y\leq k\leq y+d-1}r_k$. If $r'(y)-l'(y)\geq d+1$, then we find the empty square. That is $(x,y,d)$, where $x=l'(y)$.
We can check each $l'(k)$ and $r'(k)$ for each $y\in \{j-d+1,\dots,j\}$ using minimum queue. The data structure is a modification of the queue data structure that allows us to obtain a minimum of elements in $O(1)$. For a detailed description of the data structure, you can check for example \cite{emaxxMinQueue} or  \cite{stackoverflowMinQueue}.
Firstly, we add all elements $r_k$ for $k$ from $j-d+1$ to $j$; and we can compute $r'(j-d+1)$. After that for moving from $y-1$ to $y$, we remove an element $r_{y-1}$ from the queue and add $r_{y+d-1}$. After that all elements for computing $r'(y)$ are in the queue and we can compute it in $O(1)$. The same strategy is for $l'(y)$ but we use the maximum queue.


Let us have the following procedures for the minimum and maximum queues:
 $\textsc{InitMinQueue}()$ returns an empty minimum queue;
 $\textsc{InitMaxQueue}()$ returns an empty maximum queue; 
 $\textsc{Add}(q,u)$ adds an element $u$ to a minimum or maximum queue $q$;
 $\textsc{Remove}(q)$ removes the last element from a minimum or maximum queue $q$;
 $\textsc{Min}(q)$ returns the minimum of elements from the minimum queue $q$;
$\textsc{Max}(q)$ returns the maximum of elements from the maximum queue $q$;

The implementation is presented in Algorithm \ref{alg:fixsize2}(Appendix \ref{apx:alg-fixsize2}), and Lemma \ref{lm:fixlen2} describes the complexity.

\begin{lemma}\label{lm:fixlen2}
The query complexity of Algorithm \ref{alg:fixlen} for  $\textsc{FixedSizeFixedPointSquare}(f,i,j,d)$ subroutine is $O(d^{1.5}\cdot T_f)$ and the error probability is at most $0.1$, where $T_f$ is the query complexity of computing $f$. (See Appendix \ref{apx:fixlen2})
\end{lemma}

Secondly, we solve a more complex problem. Let us fix a size $d$ such that $1\leq d\leq n$. We want to implement a subroutine $\textsc{FixedSizeSquare}(f,d)$ that returns any empty square $(x,y,d)$ and $(NULL,NULL,NULL)$ if there is no such a square.

Let us suggest a randomized algorithm that implements the subroutine similar to the one-dimensional case. Let us uniformly randomly choose $(i,j)\in_R\{0,\dots,n-1\}\times\{0,\dots,n-1\}$; invoke $\textsc{FixedSizeFixedPointSquare}(f,i,j,d)$; and return its result.
Let us discuss the error probability of the implementation. If there is no empty square of the size $d$, then the randomized implementation always returns $(NULL,NULL,NULL)$. If there is at least one such square, then the probability of the right answer is the probability of picking a point $(i,j)$ inside an empty square of the size $d$. At the same time, we have the error probability for $\textsc{FixedSizeFixedPointSquare}(f,i,j,d)$. So, the total success probability is $p_{success}\leq\frac{0.9\cdot (d')^2}{n^2}$ if the maximal size of an empty square is $d'$ and $d\leq d'$.
We apply the Amplitude Amplification algorithm \cite{bhmt2002} for the randomized implementation and obtain error probability $0.1$. If $d=O(d')$, then the total query complexity of the algorithm is $O(\sqrt{\frac{1}{p_{success}}}\cdot d^{1.5}\cdot T_f)=O(\sqrt{\frac{n^2}{(d')^2}}\cdot d^{1.5}\cdot T_f)=O(n\cdot \sqrt{d}\cdot T_f)$.
The error probability for the amplitude amplification implementation of  $\textsc{FixedSizeSquare}(f,d)$ is $0.1$.

Finally, let us discuss the algorithm for $LSQR(f,n)$ problem.
We use exactly the same algorithm as for $LSEG$. It is Algorithm \ref{alg:lseg}, but we use  $\textsc{FixedSizeSquare}(f,d)$  instead of  $\textsc{FixedLen}(f,d)$.
Let us discuss the complexity of the algorithm.
\begin{theorem}\label{th:lsqr}
The presented algorithm solves $LSQR(f,n)$ and has $O(n^{1.5}\log n \log \log n\cdot T_f)$ query complexity and $0.1$ error probability, where $T_f$ is the query complexity of computing $f$. (See Appendix \ref{apx:lsqr})
\end{theorem}
Let us discuss the classical (deterministic or randomized) and quantum lower bounds for the problem's complexity.
\begin{theorem}\label{th:lsqr-lb}
Assume that $T_f=O(1)$, then the lower bound for query complexity for $LSQR(f,n)$ is $\Omega(n^2)$ in the classical case, and $\Omega(n)$ in the quantum case. (See Appendix \ref{apx:lsqr-lb})
\end{theorem}
So, we obtain a quantum speed up but the quantum upper and lower bounds are different.

\textbf{The Largest Empty Rectangle Problem with Fixed Width. } 
Let us consider the first type of the empty rectangle problem which is The Largest Empty Rectangle Problem with Fixed Width $LRECW(f,n,d)$.
We use ideas similar to the one-dimensional case.

Firstly, we solve a simple problem. Let us fix a column $x_1$ such that $0\leq x_1\leq n-d$; we also can compute a column $x_2=x_1+d-1$. Our goal is to search for the largest empty rectangle $(x_1,y_1,x_2,y_2)$. We define a subroutine $\textsc{FixedLeftRec}(f,x_1)$ that returns the largest empty rectangle of width $d$ with left side $x_1$. Formally, the subroutine returns an empty rectangle $(x_1,y_1,x_1+d-1,y_2)$ of the maximal possible size. If there is no such an empty rectangle, then it returns $(NULL,NULL,NULL,NULL)$. We assume that $size(NULL,NULL,NULL,NULL)=0$.

Let us consider a function $g_{x_1}:\{0,\dots,n-1\}\to\{0,1\}$ such that $g_{x_1}(i)=0$ iff $f(i,j)=0$ for all $x_1\leq j\leq x_1+d-1$. In other words, all elements of $i$-th row between $x_1$ and $x_1+d-1$ are $0$.
For computing the function, we can use Grover's search algorithm.
The solution of  $\textsc{FixedLeftRec}(f,x_1)$ is exactly solution of $LSEG(g_{x_1},n)$. So, we use Algorithm \ref{alg:lseg} for $LSEG(g_{x_1},n)$.
The final solution for the problem is searching for the rectangle $(x_1,y_1,x_1+d-1,y_2)$ with the largest size that is $\max\limits_{0\leq x_1\leq n-d} size(\textsc{FixedLeftRec}(f,x_1))$. We search for it using quantum maximum search algorithm \cite{dh96,dhhm2006}. We invoke nested Grover search algorithms ($\textsc{FixedLeftRec}$ inside maximum). That is why we should use the bounded-error input version of the Grover search algorithm that is presented in \cite{hmw2003,abikkpssv2020,abikkpssv2023}.
The complexity of the algorithm is presented in the following theorem.
\begin{theorem}\label{th:lrecw}
The presented algorithm solves $LRECW(f,n,d)$ and has $O(n\sqrt{d}\log n\cdot T_f)$ query complexity and $0.1$ error probability, where $T_f$ is the query complexity of computing $f$. (See Appendix \ref{apx:lrecw})
\end{theorem}
Let us discuss the classical (deterministic or randomized) and quantum lower bounds for the problem's complexity.
\begin{theorem}\label{th:lrecw-lb}
Assume that $T_f=O(1)$, then the lower bound for query complexity for $LRECW(f,n,d)$ is $\Omega(nd)$ in the classical case, and $\Omega(\sqrt{nd})$ in the quantum case. (See Appendix \ref{apx:lrecw-lb})
\end{theorem}
So, we obtain a quantum speed up but the quantum upper and lower bounds are different.

\textbf{The Largest Empty Rectangle Problem for Rectangle Empty Areas. }
Let us solve a simple version of the problem that is the problem for a fixed point $(i,j)$. We want to check the size of an empty rectangle that contains $(i,j)$ for $0\leq i\leq n-1, 0\leq j\leq n-1$.
We use the following idea:

 {\bf Step 1}. If $f(i,j)=1$, then the size is $0$. Otherwise, go to Step 2.

 {\bf Step 2}. We search for the closest $1$-value to the left. That is $x_1\gets \textsc{LastOne}(f(i),0,j)$.  We search for the closest $1$-value to the right. That is $x_2\gets \textsc{FirstOne}(f(i),j,n-1)$.  We search for the closest $1$-value to the up direction. 
That is $y_1\gets \textsc{LastOne}(f(*,j),0,i)$.  Here, $f(*,j):\{0,\dots,n-1\}\to\{0,1\}$ such that $f(*,j)(i)=f(i,j)$. We search for the closest $1$-value to the down direction.  That is $y_2\gets \textsc{FirstOne}(f(*,j),i,n-1)$. The result empty rectangle is $(x_1+1,y_1+1,x_2-1,y_2-1)$.

The implementation of the idea is presented in Algorithm \ref{alg:recemptyarea} (Appendix \ref{apx:alg-recemptyarea}) as $\textsc{FixedPointRecArea}(f,i,j)$; and the complexity is discussed in Lemma \ref{lm:fixlen3}.

\begin{lemma}\label{lm:fixlen3}
The query complexity of Algorithm \ref{alg:recemptyarea} for  $\textsc{FixedPointRecArea}(f,i,j)$ subroutine is $O(\sqrt{n}\cdot T_f)$ and the error probability is at most $0.1$, where $T_f$ is the query complexity of computing $f$. (See Appendix \ref{apx:fixlen3})
\end{lemma}

Let us assume that $size(NULL,NULL,NULL,NULL)=0$.
Let us consider $(x_1,y_1,x_2,y_2)$ is such that $size(x_1,y_1,x_2,y_2)=\max\limits_{0\leq i\leq n-1, 0\leq j\leq n-1}\textsc{FixedPointRecArea}(f,i,j)$.
We search for it using quantum maximum search algorithm \cite{dh96,dhhm2006}. Here, we also use the bounded-error input version of the Grover search algorithm \cite{hmw2003,abikkpssv2020,abikkpssv2023}.
The complexity of the algorithm is presented in the following theorem.
\begin{theorem}\label{th:lrec22}
The presented algorithm solves $LREC2(f,n)$ and has $O(n^{1.5}\cdot T_f)$ query complexity and $0.1$ error probability, where $T_f$ is the query complexity of computing $f$. (See Appendix \ref{apx:lrec22})
\end{theorem}

Let us discuss the classical (deterministic or randomized) and quantum lower bounds for the problem's complexity.
\begin{theorem}\label{th:lrec2}
Assume that $T_f=O(1)$, then the lower bound for query complexity for $LREC2(f,n)$ is $\Omega(n^2)$ in the classical case, and $\Omega(n)$ in the quantum case. (See Appendix \ref{apx:lrec2})
\end{theorem}

So, we obtain a quantum speed up but the quantum upper and lower bounds are different.
\textbf{The Largest Empty Rectangle Problem. }
Unfortunately, we cannot present a quantum algorithm for The Largest Empty Rectangle Problem that shows quantum speed-up. At the same time, we can present the lower bound.
\begin{theorem}\label{th:lrec2-lb}
Assume that $T_f=O(1)$, then the lower bound for query complexity for $LREC(f,n,h,w)$ is $\Omega(n^2)$ in the classical case, and $\Omega(n)$ in the quantum case. (See Appendix \ref{apx:lrec2-lb})
\end{theorem}

\section{Conclusion}\label{sec:conclusion}
In this paper, we discuss the largest area search problems for one and two-dimensional cases. In the one-dimensional case, we presented a quantum algorithm with $O(\sqrt{n}\log n\log\log n)$ query complexity. We show that it is equal to the lower bound up to the log factor. The quantum lower bound is $\Omega(\sqrt{n})$. At the same time, we reach almost quantum quadratic speed-up. Any classical algorithm has query complexity of at least $\Omega(n)$.

In two dimensional case, we reach not so significant, but still important speed-up. We presented quantum solutions for $LREC2$, $LSQR$, and $LRECW$ problems. The query complexity is $O(n^{1.5})$ for $LREC2$, is $O(n^{1.5}\log n\log \log n)$ for $LSQR$, and $O(n\sqrt{d}\log n\log \log n)$ for $LRECW$.
At the same time, the lower bounds for classical solutions are $\Omega(n^2)$, $\Omega(n^2)$, and $\Omega(nd)$, respectively. So, we obtain quantum polynomial speed-up.
At the same time, the presented quantum lower bound is $\Omega(n)$, and it is far from the upper bounds. 
The open questions are the following.
\begin{itemize}
    \item 
Can we improve lower bounds or upper bounds for two-dimensional problems?
    \item 
Can we develop a quantum solution for the general $LREC$ problem that shows quantum speed up? Maybe, we should improve the quantum lower bound.
    \item 
Can we improve the algorithm for the one-dimensional case and remove the log factors?
\end{itemize}

\textbf{Acknowledgements}
The research has been supported by Russian Science Foundation Grant 25-11-00366, \url{https://rscf.ru/project/25-11-00366/}.

%
%
%

\bibliographystyle{unsrt}
\bibliography{tcs}
\newpage
\appendix
\section{Algorithm \ref{alg:fixlen}}\label{apx:alg-fixlen}
 \begin{algorithm}[H]
    \caption{$\textsc{FixedLenFixedPoint}(f,i,d)$ subroutine implementation} \label{alg:fixlen}
    \begin{algorithmic}
        \State $result \gets (NULL,NULL)$
        \If{$f(i)=0$}\Comment{Step 1.}
        \State $r\gets\textsc{FirstOne}(f,i,min(n-1,i+d-1))$\Comment{Step 2.}
        \If{$r = NULL$}
             \State $r\gets n-1$ 
          \EndIf 
          \State $\ell\gets\textsc{LastOne}(f,max(0,r-d+1),r)$
          \If{$\ell=NULL$}
          \State $\ell\gets r-d+1$
          \EndIf
          \If{$r-\ell+1<d$}
            \State $result\gets(NULL,NULL)$
            \Else
            \State $result\gets(\ell,r)$
          \EndIf
        \EndIf
        \State \Return{$result$}
    \end{algorithmic}
\end{algorithm}

\section{Algorithm \ref{alg:lseg}}\label{apx:alg-lseg}
  \begin{algorithm}[H]
    \caption{Solution of the $LSEG(f,n)$ problem} \label{alg:lseg}
    \begin{algorithmic}
        \State $d\gets 1$
        \State $seg \gets \textsc{FixedLen}(f,d\cdot 2)$\Comment{repeat $2\log_2\log_2 n$ times}
        \While{$seg\neq (NULL,NULL)$}
          \State $d\gets d\cdot 2$
          \State $seg \gets \textsc{FixedLen}(f,d\cdot 2)$\Comment{repeat $2\log_2\log_2 n$ times}
        \EndWhile
        \State $left\gets d$, $right\gets 2\dot d $
        \While{$left<right$}
        \State $middle = \lfloor(left+right)/2\rfloor$
        \If{$\textsc{FixedLen}(f, middle)=(NULL,NULL)$}\Comment{repeat $2\log_2\log_2 n$ times}
        \State $right\gets middle $
        \Else 
        \State $left\gets middle+1$
        \EndIf
        \EndWhile
        \State $result\gets \textsc{FixedLen}(f, left-1)$\Comment{repeat $2\log_2\log_2 n$ times}
        \State \Return{$result$}
    \end{algorithmic}
\end{algorithm}

\section{Proof of Theorem \ref{th:compl-lseg}}\label{apx:compl-lseg}
\textbf{Theorem \ref{th:compl-lseg}. }{\em
Algorithm \ref{alg:lseg} solves $LSEG(f,n)$ and has $O(\sqrt{n}\log n\cdot \log \log n\cdot T_f)$ query complexity and $O\left(\frac{1}{\log n}\right)$ error probability.
}

\Beginproof On the Step 1, we invoke $\textsc{FixedLen}$ subroutine $O(\log d'\log\log n)$ times. Here, we can be sure that $d$ for $\textsc{FixedLen}(f,d)$ is such that $d\leq 2\cdot d'$. So, $d=O(d')$.  The query complexity of the subroutine is $O(\sqrt{n}\cdot T_f)$. Therefore, the query complexity of Step 1 is $O(\sqrt{n}\log d'\log\log n\cdot T_f)=O(\sqrt{n}\log n\log\log n\cdot T_f)$.
The query complexity of Step 2 is the same. The total query complexity is $O(\sqrt{n}\log n\log\log n\cdot T_f)$.

We have $2\log_2\log_2 n$ repetitions of $\textsc{FixedLen}(f, middle)$. Due to ``one-side'' error behaver of the $\textsc{FixedLen}$ function, these invocations have an error only if all of them have errors. So, the error probability is at most $\frac{1}{2^{2log_2log_2 n}}=\frac{1}{(\log_2 n)^2}$. The algorithm invokes this subroutine $2\log_2 n$ times because of the Binary lifting (Step 1) and the Binary search (Step 2). The total error probability is
$1-\left(1-\frac{1}{(\log_2 n)^2}\right)^{2\log_2n}=O\left(\frac{1}{\log n}\right)$.
\Endproof

\section{Proof of Theorem \ref{th:compl-lseg2}}\label{apx:compl-lseg2}
\textbf{Theorem \ref{th:compl-lseg2}. }{\em
Assume that $T_f=O(1)$, then the lower bound for the query complexity for $LSEG(f,n)$ is $\Omega(n)$ in the classical case, and $\Omega(\sqrt{n})$ in the quantum case. 
}

\Beginproof
Let us consider the $LSEG(f,n)$ problem where possible inputs are $f^k$ for $k\in\{n/2+1,\dots,n-1\}$. Here,  $f^k(i)=0$ for all $1\leq i\neq k$; and $f^k(0)=f^k(k)=1$.
So, we can say that the length of the largest segment for $f^k$ is $k-1$.
Therefore, the problem is at least as hard as an unstructured search for ones among elements with indices from  $\{n/2+1,\dots,n-1\}$. The lower bound for the unstructured search is $\Omega(n)$ in the classical (deterministic or randomized) case, and $\Omega(\sqrt{n})$ in the quantum case \cite{bbbv1997}.
Assume that there is a solution of  $LSEG(f,n)$ with $o(\sqrt{n})$ query complexity. Then, we can use this solution for solving the presented the unstructured search problem. So, we obtain a contradiction.
\Endproof
\section{Algorithm \ref{alg:fixlenszbt}}\label{apx:alg-fixlenszbt}
\begin{algorithm}[H]
    \caption{$\textsc{FixedLenFixedPointSZBT}(f,i,d)$ subroutine implementation} \label{alg:fixlenszbt}
    \begin{algorithmic}
        \State $result \gets (NULL,NULL)$
        \If{$f(i)=0$}\Comment{Step 1.}
        \State $r\gets\textsc{FirstOne}(f',i,min(n,i+d-1))$\Comment{Step 2.}
        \If{$r = NULL$}
             \State $r\gets n$ 
          \EndIf 
          \If{$f(r)=2$}
          	\State $\ell\gets\textsc{LastOne}(f,r-d+1,r)$
          	\If{$\ell\neq NULL$ and $f(\ell)\neq 1$}
          		\State $result\gets(\ell,r)$
            \Else
          \EndIf
          \EndIf
        \EndIf
        \State \Return{$result$}
    \end{algorithmic}
\end{algorithm}
\section{Algorithm \ref{alg:fixsize2}}\label{apx:alg-fixsize2}
    \begin{algorithm}[H]
    \caption{$\textsc{FixedSizeFixedPointSquare}(f,i,j,d)$ subroutine implementation (Idea 2)} \label{alg:fixsize2}
    \begin{algorithmic}
        \State $result \gets (NULL,NULL,NULL)$
        \If{$f(i,j)=0$}\Comment{Step 1.}
        \For{$k\in\{i-d+1,\dots, i+d-1\}$}\Comment{Step 2.}
        \State  $l_k\gets\textsc{LastOne}(f(k),j,j-d+1)$,  $r_k\gets\textsc{FirstOne}(f(k),j,j+d-1)$\Comment{repeat $2\log_2 d$ times}
        \If{$l_k=NULL$}
        \State  $l_k\gets j-d$
        \EndIf
         \If{$r_k=NULL$}
        \State  $r_k\gets j+d$
        \EndIf
        \EndFor
        
        \State $minq\gets\textsc{InitMinQueue}(),maxq\gets\textsc{InitMaxQueue}()$\Comment{Step 3.}
        \For{$k\in\{j-d+1,\dots,j\}$}
        \State $\textsc{Add}(minq, r_k)$, $\textsc{Add}(maxq, l_k)$
        \EndFor
        \If{$\textsc{Min}(minq)-\textsc{Max}(maxq)\geq d+1$}
        \State $result\gets(\textsc{Max}(maxq),j-d+1, d)$
        \Else
        \For{$y\in\{j-d+2,\dots,j\}$}
        \State $\textsc{Remove}(minq, r_{y-1})$, $\textsc{Remove}(maxq, l_{y-1})$
        \State $\textsc{Add}(minq, r_{y+d-1})$, $\textsc{Add}(maxq, l_{y+d-1})$
        \If{$\textsc{Min}(minq)-\textsc{Max}(maxq)\geq d+1$}
        \State $result\gets(\textsc{Max}(maxq), y, d)$
        \State stop the for-loop.
        \EndIf
        \EndFor
        \EndIf
        \EndIf
        \State \Return{$result$}
    \end{algorithmic}
\end{algorithm}
\section{Proof of Lemma \ref{lm:fixlen2}}\label{apx:fixlen2}
\textbf{Lemma \ref{lm:fixlen2}. }{\em
The query complexity of Algorithm \ref{alg:fixlen} for  $\textsc{FixedSizeFixedPointSquare}(f,i,j,d)$ subroutine is $O(d^{1.5}\cdot T_f)$ and the error probability is at most $0.1$, where $T_f$ is the query complexity of computing $f$.
}

\Beginproof
The query complexity of Step 1 is constant. Let us discuss the query complexity of Step 2. 

The query complexity for $\textsc{LastOne}(f(k),j,j-d+1)$ and $\textsc{FirstOne}(f(k),j,j+d-1)$ is $O(\sqrt{d}\cdot T_f)$. We invoke these subroutines $2d$ times. So, the total complexity is $O(d^{1.5}\cdot T_f)$.  Each invocation can have an error and the total error accumulates very fast. According to \cite{k2014}, it can be converted to the algorithm that does the same action, has the same complexity but has a total error $0.1$.

The query complexity for Step 3 is $O(d)$ because we added $O(d)$ elements to a queue and return the minimum and the maximum in $O(1)$ for each position of the ``window''. The complexity of adding to a queue is $O(1)$. At the same time, the complexity of removing an element is $O(d)$ for one of the invocations and $O(1)$ for all others. Therefore, we can say that the total complexity of all removing is also $O(d)$. See section ``Finding the minimum for all subarrays of fixed length'' from \cite{emaxxMinQueue} or \cite{stackoverflowMinQueue} for more details.

Step 3 is deterministic and does not have an error. So, the total error probability is at most $0.1$ and the query complexity is $O(d^{1.5}\cdot T_f)$.
\Endproof

\section{Proof of Theorem \ref{th:lsqr}}\label{apx:lsqr}
\textbf{Theorem \ref{th:lsqr}. }{\em
The presented algorithm solves $LSQR(f,n)$ and has $O(n^{1.5}\log n \log \log n\cdot T_f)$ query complexity and $0.1$ error probability, where $T_f$ is the query complexity of computing $f$. 
}

\Beginproof
The proof is similar to the proof of Theorem \ref{th:compl-lseg}.
The query complexity of Step 1 is $O(n\sqrt{d'}\log d'\log\log n\cdot T_f)=O(n^{1.5}\log n\log\log n\cdot T_f)$.
The query complexity of Step 2 is the same. The total query complexity is $O(n^{1.5}\log n\log\log n\cdot T_f)$.

$2\log_2\log_2 n$ repetitions of $\textsc{FixedSizeSquare}(f, middle)$ have an error only if all invocations have an error. So, the error probability is at most $\frac{1}{2^{2log_2log_2 n}}=\frac{1}{(\log_2 n)^2}$. The algorithm invokes this subroutine $2\log_2 n$ times because of the Binary lifting (Step 1) and the Binary Search (Step 2). The total error probability is
$1-\left(1-\frac{1}{(\log_2 n)^2}\right)^{2\log_2n}=O\left(\frac{1}{\log n}\right).$
\Endproof

\section{Proof of Theorem \ref{th:lsqr-lb}}\label{apx:lsqr-lb}
\textbf{Theorem \ref{th:lsqr-lb}. }{\em
Assume that $T_f=O(1)$, then the lower bound for query complexity for $LSQR(f,n)$ is $\Omega(n^2)$ in the classical case, and $\Omega(n)$ in the quantum case.
}

\Beginproof
Let us consider the $LSQR(f,n)$ problem and possible inputs are $f^{k,t}$ for $(k,t)\in\{n/2+1,\dots,n-1\}\times\{n/2+1,\dots,n-1\}$. Here,  $f^{k,t}(i,j)=0$ for all $(i,j)\in\{0,\dots, n-1\}\times\{0,\dots,n-1\}\backslash\{(0,0),(k,t)\}$; and $f^{k,t}(0,0)=f^{k,t}(k,t)=1$.
So, we can say, that the size of the largest square for $f^{k,t}$ is $min(k-1,t-1)$.

Assume that we can solve $LSQR(f,n)$ using $o(n^2)$ query complexity classically or $o(n)$ query complexity quantumly. In that case, we find the position of $1$ in that time. If we find the largest empty square, then we determine a row or a column of the $1$. After that, we can find its index using the Linear Search in a line in the classical case (in $O(n)$) or Grover search in the quantum case (in $O(\sqrt{n}$). Therefore, we can solve the unstructured search problem among $\Omega(n^2)$ bits faster than the existing lower bound \cite{bbbv1997}.  This is a contradiction.
\Endproof

\section{Proof of Theorem \ref{th:lrecw}}\label{apx:lrecw}
\textbf{Theorem \ref{th:lrecw}. }{\em
The presented algorithm solves $LRECW(f,n,d)$ and has $O(n\sqrt{d}\log n\cdot T_f)$ query complexity and $0.1$ error probability, where $T_f$ is the query complexity of computing $f$.
}

\Beginproof
 Query complexity for computing $g_{x_1}$ is $T_{g_{x_1}}=O(\sqrt{d}\cdot T_f)$ and the error probability is at most $0.1$ because of Grover's search algorithm's properties.
 
 For computing $\textsc{FixedLeftRec}(f,x_1)$, we use algorithm for $LSEG(g_{x_1},n)$ and Grover's search algorithm for computing $g_{x_1}$. Due to Theorem \ref{th:compl-lseg}, query complexity for 
the problem is $O(\sqrt{n}\log n\log\log n\cdot T_{g_{x_1}})=O(\sqrt{nd}\log n\log\log n\cdot T_f)$.

Due to \cite{dh96,dhhm2006}, complexity of maximum searching is $O(\sqrt{n})$.
The final query complexity is $O(\sqrt{n}\cdot \log n\log\log n\cdot T_f)=O(n\sqrt{d}\log n\log\log n\cdot T_f)$. The error probability is at most $0.1$ due to \cite{hmw2003,abikkpssv2020,abikkpssv2023}.
\Endproof
\section{Proof of Theorem \ref{th:lrecw-lb}}\label{apx:lrecw-lb}
\textbf{Theorem \ref{th:lrecw-lb}. }{\em
Assume that $T_f=O(1)$, then the lower bound for query complexity for $LRECW(f,n,d)$ is $\Omega(nd)$ in the classical case, and $\Omega(\sqrt{nd})$ in the quantum case. 
}

\Beginproof
Let us consider the $LRECW(f,n,d)$ problem and possible inputs are $f^{k,t}$ for $(k,t)\in\{0,\dots,d-1\}\times\{n/2+1,\dots,n-1\}$. Here,  $f^{k,t}(i,j)=0$ for all $(i,j)\in\{0,\dots, d-1\}\times\{0,\dots,n-1\}\backslash\{(0,0),(k,t)\}$;  $f^{k,t}(0,0)=f^{k,t}(k,t)=1$; and $f^{k,t}(i,j)=1$ for all $d\leq i\neq n-1$, $0\leq j\neq n-1$.
So, we can say that the size of the empty rectangle for $f^{k,t}$ is $t-1$.

Assume that we can solve $LRECW(f,n,d)$ using $o(nd)$ query complexity classically or $o(\sqrt{nd})$ query complexity quantumly. In that case, we find the position of $1$ in that time. If we find the largest empty rectangle, then we determine a row of the $1$. After that, we can find its index using the Linear Search in the row in the classical case (in $O(n)$) or Grover search in the quantum case (in $O(\sqrt{n}$). Therefore, we can solve the unstructured search problem among $\Omega(nd)$ bits faster than the existing lower bound \cite{bbbv1997}.  This is a contradiction.
\Endproof
\section{Algorithm \ref{alg:recemptyarea}}\label{apx:alg-recemptyarea}
 \begin{algorithm}[H]
    \caption{$\textsc{FixedPointRecArea}(f,i,j)$ subroutine implementation} \label{alg:recemptyarea}
    \begin{algorithmic}
        \State $result \gets (NULL,NULL,NULL,NULL)$
        \If{$f(i,j)=0$}\Comment{Step 1.}
        \State$x_1\gets \textsc{LastOne}(f(i),0,j)$, $x_2\gets \textsc{FirstOne}(f(i),j,n-1)$\Comment{Step 2.}
        \State$y_1\gets \textsc{LastOne}(f(*,j),0,i)$, $y_2\gets \textsc{FirstOne}(f(*,j),i,n-1)$
        \State$result\gets(x_1,y_1,x_2,y_2)$
        \EndIf
        \State \Return{$result$}
    \end{algorithmic}
\end{algorithm}

\section{Proof of Lemma \ref{lm:fixlen3}}\label{apx:fixlen3}
\textbf{Lemma \ref{lm:fixlen3}. }{\em
The query complexity of Algorithm \ref{alg:recemptyarea} for  $\textsc{FixedPointRecArea}(f,i,j)$ subroutine is $O(\sqrt{n}\cdot T_f)$ and the error probability is at most $0.1$, where $T_f$ is the query complexity of computing $f$. 
}

\Beginproof
Query complexity of $\textsc{LastOne}$ and $\textsc{FirstOne}$ is $O(\sqrt{n}\cdot T_f)$ and the error probability is $0.1$.  We can archive the smaller error probability with a constant number of repetitions. That allows us to claim that the error probability for the algorithm is $0.1$ for the same query complexity.
\Endproof

\section{Proof of Theorem \ref{th:lrec22}}\label{apx:lrec22}
\textbf{Theorem \ref{th:lrec22}. }{\em
The presented algorithm solves $LREC2(f,n)$ and has $O(n^{1.5}\cdot T_f)$ query complexity and $0.1$ error probability, where $T_f$ is the query complexity of computing $f$.
}

\Beginproof
 Query complexity for \textsc{FixedPointRecArea} is $O(\sqrt{n}\cdot T_f)$ due to the previous lemma.
Due to \cite{dh96,dhhm2006}, complexity of maximum searching is $O(\sqrt{n^2})=O(n)$.
The final query complexity is $O(n^{1.5}\cdot T_f)$. The error probability is at most $0.1$ due to \cite{hmw2003,abikkpssv2020,abikkpssv2023}.
\Endproof

\section{Proof of Theorem \ref{th:lrec2}}\label{apx:lrec2}
\textbf{Theorem \ref{th:lrec2}. }{\em
Assume that $T_f=O(1)$, then the lower bound for query complexity for $LREC2(f,n)$ is $\Omega(n^2)$ in the classical case, and $\Omega(n)$ in the quantum case. 
}

\Beginproof
Let us consider the $LREC2(f,n,d)$ problem and possible inputs are $f^{k,t}$ for $(k,t)\in\{0,\dots,n-1\}\times\{0,\dots,n-1\}\cup\{(-1,-1)\}$. Here,  $f^{k,t}(i,j)=1$ for all $(i,j)\in\{0,\dots, d-1\}\times\{0,\dots,n-1\}\backslash\{(k,t)\}$;  $f^{k,t}(k,t)=0$. If $(k,t)=(-1,-1)$, then all $f^{-1,-1}(i,j)=1$.
So, we can say that the in the case of $f^{-1,-1}$ the result is $(NULL,NULL,NULL,NULL)$, and in the other case, it is $(k,t,k,t)$.

Assume that we can solve $LREC2(f,n,)$ using $o(n^2)$ query complexity classically or $o(n)$ query complexity quantumly. In that case, we find the position of $0$ in that time. Therefore, we can solve the unstructured search problem among $\Omega(n^2)$ bits faster than the existing lower bound \cite{bbbv1997}.  This is a contradiction.
\Endproof

\section{Proof of Theorem \ref{th:lrec2-lb}}\label{apx:lrec2-lb}
\textbf{Theorem \ref{th:lrec2-lb}. }{\em
Assume that $T_f=O(1)$, then the lower bound for query complexity for $LREC(f,n,h,w)$ is $\Omega(n^2)$ in the classical case, and $\Omega(n)$ in the quantum case. 
}

\Beginproof
Let us consider the $LREC(f,n,h,w)$ and possible inputs are $f^{k,t}$ for $k,t\in\{0,\dots,n-1\}$ or $f^{-1,-1}$. Here,  $f^{k,t}(i,j)=0$ for all $(i,j)\neq (k,t)$; and $f^{k,t}(k,t)=1$. The function $f^{-1,-1}$ is such that $f^{-1,-1}(i,j)=0$ for all $(i,j)$.

If $f=f^{-1,-1}$, then the largest empty rectangle has size $n^2$. In other cases, it is less than $n^2$. Therefore, if we can solve $LREC(f,n,h,w)$, then we can distinguish between $f^{-1,-1}$ case and $f^{k,t}$ case. So, the $LREC(f,n,h,w)$ is as hard as a problem of distinguishing these two cases.
The distincting problem is as hard as searching for $1$ among the $n\times n$-matrix. 
Hence, the lower bound for $LREC(f,n,h,w)$ is the same as the lower bound for unstructured search among $n^2$ elements.
That is $\Omega(n^2)$ in the classical (deterministic or randomized) case, and $\Omega(\sqrt{n^2})=\Omega(n)$ in the quantum case.
\Endproof

\end{document}